\documentclass[useAMS,usenatbib]{mn2e}
\usepackage{journal_names}
\usepackage{graphicx,times}
\usepackage{amsmath}
\usepackage[T1]{fontenc}
\usepackage{aecompl}
\usepackage{subfig}
\usepackage{tabularx}

\title[On How to Extend the NIR Tully-Fisher Relation to be Truly All-Sky]{On How to Extend the NIR Tully-Fisher Relation to be Truly All-Sky}
\author[K. Said, R. C. Kraan-Korteweg and T. H. Jarrett]{K. Said\thanks{E-mail:
khaled@ast.uct.ac.za}, R. C. Kraan-Korteweg and T. H. Jarrett\\
Astrophysics, Cosmology and Gravity Centre (ACGC), Astronomy Department, 
University of Cape Town, Private Bag X3, Rondebosch, 7701, South Africa}
\begin{document}

\date{Accepted 2014 November 25. Received 2014 November 14; in original form 2014 July 09}

\pagerange{\pageref{firstpage}--\pageref{lastpage}} \pubyear{2014}

\maketitle

\label{firstpage}

\begin{abstract}
Dust extinction and stellar confusion by the Milky Way reduce the efficiency of detecting galaxies at low Galactic latitudes, creating the so-called Zone of Avoidance. This stands as a stumbling block in charting the distribution of galaxies and cosmic flow fields, and therewith our understanding of the local dynamics in the Universe (CMB dipole, convergence radius of bulk flows). For instance, ZoA galaxies are generally excluded from the whole-sky Tully-Fisher Surveys ($|b| \leq 5^\circ$) even if catalogued. We show here that by fine-tuning the near-infrared TF relation, there is no reason not to extend peculiar velocity surveys deeper into the ZoA. Accurate axial ratios ($b/a$) are crucial to both the TF sample selection and the resulting TF distances. We simulate the effect of dust extinction on the geometrical properties of galaxies. As expected, galaxies appear rounder with increasing obscuration level, even affecting existing TF samples. We derive correction models and demonstrate that we can reliably reproduce the intrinsic axial ratio from the observed value up to extinction level of about $A_J\simeq3$ mag ($A_V\sim11$ mag), we also recover a fair fraction of galaxies that otherwise would fall out of an uncorrected inclination limited galaxy sample. We present a re-calibration of the 2MTF relation in the NIR $J$, $H$, and $K_s$-bands for isophotal rather than total magnitudes, using their same calibration sample. Both TF relations exhibit similar scatter at high Galactic latitudes. However, the isophotal TF relation results in a significant improvement in the scatter for galaxies in the ZoA, and low surface brightness galaxies in general, because isophotal apertures are more robust in the face of significant stellar confusion. 
\end{abstract}

\begin{keywords}
galaxies: distances and peculiar velocities -- zone of avoidance: dust extinction and stellar confusion -- tully-fisher -- isophotal magnitudes -- axial ratio
\end{keywords}

\section{Introduction}
The difference between the observed velocity of a galaxy and the smooth Hubble flow is known as peculiar
velocity. Because this peculiar velocity arises as a result of gravity, it can be used as a tracer of the matter distribution, both luminous and dark. Of key importance is the motion of the Local Group (LG) relative to the Cosmic Microwave Background (CMB) \citep{1969Natur.222..971C, 1971Natur.231..516H}. This motion is caused by the gravitational influence of density inhomogeneities in the volume surrounding the LG. Determination of the direction of this peculiar motion requires a census of galaxies across the whole sky.

Dust extinction and stellar confusion by the Milky Way create the so-called Zone of Avoidance (ZoA). Due to the lack of data in the ZoA, current redshift and peculiar velocity whole-sky surveys exclude the ZoA from their analysis (e.g. \citealt{2012ApJS..199...26H,2008AJ....135.1738M}). Many attempts have been made to artificially fill or recreate the ZoA to measure the so-called \textit{``clustering dipole''} (\citealt{2011ApJ...741...31B}). For example \citet{2006MNRAS.368.1515E,2006MNRAS.373...45E} reconstructed the local Universe using two different methods to repopulate the ZoA, but one of the main components of uncertainty in the resulting dipole remains the incomplete mapping of the ZoA \citep{2008MNRAS.386.2221L}. Past surveys in the ZoA revealed large-scale structures like the Great Attractor (GA) \citep{1988ApJ...326...19L,1999A&A...352...39W}, Perseus-Pisces (PP) \citep{1980MNRAS.193..353E,1982AJ.....87.1355G,1984A&A...136..178F,1987A&A...184...43H}, the Local Void (LV) \citep{1987ang..book.....T,2008glv..book...13K}, the Ophiuchus Cluster \citep{1981PASJ...33...57W,2000ASPC..218..187W,2005ASPC..329..189W,2002ApJ...580..774E} and very recent discovery of a hidden massive supercluster in Vela/Puppis (Kraan-Korteweg et al. in prep.). A genuine whole-sky peculiar velocity survey will significantly improve our knowledge of the dynamics of the LG, cosmic flow fields, and the observed dipole in the CMB. 
 
One of the largest TF surveys available to date is the $I$-band (SFI++) survey \citep{2007ApJS..172..599S}, which uses both cluster and field galaxies. Because of its selection in the $I$-band, it is heavily affected by dust extinction and incomplete across the Galactic plane ($|b|<15^\circ$). The Two Micron All-Sky Survey (2MASS; \citealt{2006AJ....131.1163S}) Tully-Fisher Survey (2MTF; \citealt{2008AJ....135.1738M,2013MNRAS.432.1178H,2014MNRAS.443.1044M,2014arXiv1409.0287H}) provides a complete Tully-Fisher analysis ($i.e.$ distance and peculiar velocity) of all bright inclined spirals in the 2MASS Redshift Survey (2MRS; \citealt{2012ApJS..199...26H}). The use of 2MASS near infrared (NIR) $J$, $H$, and $K_s$ bands minimizes the ZoA as the effects of foreground extinction are reduced compared to the optical bands; nevertheless a not insignificant part of the sky remains unsampled by 2MTF ($|b|<5^\circ$ and $|b|<15^\circ$ for $|l|<30^\circ$) because the selection of 2MTF is based on 2MRS which contains only galaxies within $|b|\geq5^\circ$, and $|b|\geq8^\circ$ toward the Galactic bulge.

  Systematic HI surveys in the southern ZoA \citep{2000AJ....119.2686H,2002A&A...391..887K,2005IAUS..216..203K,2005RvMA...18...48K,2008glv..book...13K} have opened an elegant way to fill in this region because these kind of surveys do not suffer from foreground extinction or source confusion that plague optical/NIR imaging. In the near future, surveys like the Widefield ASKAP L-band Legacy All-sky Blind surveY (WALLABY)\footnote{http://www.atnf.csiro.au/research/WALLABY/proposal.html} and its sister in the northern hemisphere the Westerbork Northern Sky HI Survey (WNSHS)\footnote{http://www.astron.nl/~jozsa/wnshs/} will provide an HI survey of the whole sky \citep{2012MNRAS.426.3385D} and this will dramatically improve the situation. In combination with the ongoing deep and well resolved  NIR surveys UKIDSS and VISTA, NIR TF analysis will be applied. The method presented here will allow the use of these forthcoming data to extend the flow fields studies into the ZoA.\\

Since the discovery of the correlation between rotational velocity and absolute magnitude of spiral galaxies \citep*{1977A&A....54..661T}, it has been widely used in cosmography and peculiar velocity surveys (e.g., \citealt{1992ApJ...395...75H,1993ApJ...409...14M,1996ApJ...468L...5D,2007A&A...465...71T,2012AN....333..436C,2013AJ....146...69C,2014Natur.513...71T}).  Rotational velocities of spiral galaxies (distance-independent) can be measured using the HI line-width or the optical rotation curve (ORCs) and the absolute magnitudes (distance-dependent) derived from photometry.

 In order to have a reliable TF survey, the first step is to construct a global-unbiased TF template relation. As a secondary distance indicator, the TF template relation should be obtained from a sample of galaxies with known distances. This TF relation can then be used as the calibrated relation between the absolute magnitude and rotational velocity. Raw data and corrections used in the measurement of TF distances and peculiar velocities should be consistent with that used in the derivation of the TF relation.

In the last few decades, many TF template relations have been derived using either different samples or methodology. Shortly after the inception of the TF relation, \citet{1979ApJ...229....1A} calibrated the TF relation in the NIR $H$ (1.6$\mu m$) for spiral galaxies in clusters. The NIR suffers less from dust extinction and is more sensitive to stellar mass.  They found that in the infrared the scatter is lower than that in the $B$-band and the slope is much steeper. Based on Cepheid distances to 21 galaxies, \citet{2000ASPC..201..129S} derived the TF relation in the $B$, $V$, $R$, $I$, and $H$-bands. These relations were used to derive a value of $H_0$. At the same time, \citet{2000ApJ...533..744T}, \citet{2000ApJ...533..781R} performed a calibration of the $B$, $R$, $I$, and $K$ relations. To avoid the Malmquist bias, they used the inverse method to fit the TF relation \citep*{1988ApJ...331..620K}.

 \citet{2000ASPC..218..111B} used 2MASS $J$, $H$, and $K_s$-bands to derive TF relations in the ZoA. They used five different types of magnitudes from 2MASX \citep{2000AJ....119.2498J}. They conclude that the scatter was lowest for the isophotal $K_s$-fiducial 20 mag/arcsec$^2$ TF relation. \citet{2001PhDT.......242M} derived $R$, $I$, $H$, and $K_s$-band TF relations based on Cepheid distances. He used both total and isophotal magnitudes with the 20\% HI-line-width (the width measured at 20\% of the peak flux). His results indicate that isophotal $H$, $K_s$ magnitudes and total $I$  magnitudes relations yield comparable distances. These relations have a scatter of $\sim$ 0\fm2. These two previous studies prove that isophotal magnitude works equally well in and out of the ZoA.
 
 \citet{2006ApJ...653..861M} followed the same procedure described by \citet{1997AJ....113...22G} but with a larger sample to derive a new $I$-band TF relation. Their sample consists of 807 galaxies in the vicinity of 31 clusters and groups. This template relation has been used by \citet{2007ApJS..172..599S} to derive peculiar velocities of galaxies in the SFI++ catalog, which contains 4861 field and cluster spiral galaxies. Later, \citet{2008AJ....135.1738M} used similar procedures with a slightly larger sample to derive the absolute calibration of the $J$, $H$, and $K_s$-bands relations. These relations are the first step towards the 2MTF.
 
 Recently, two independent groups have extended the TF to mid-infrared wavelengths. \citet{2013ApJ...765...94S} use Spitzer ($3.6 \mu m$) photometry for 213 cluster galaxies to derive the TF relation. \citet{2013ApJ...771...88L} use the mid-infrared ($3.4 \mu m$) $W1$-band of the Wide-field Infrared Survey Explorer ($WISE$) satellite with a larger sample of 568 field and cluster galaxies drawn from the 2MTF catalog to derive the TF relation. While \citet{2013ApJ...765...94S} adopted the inverse fitting scheme, \citet{2013ApJ...771...88L} used a bivariate scheme, where errors in both $x$ and $y$ are taken into consideration. More recently, \cite{2014ApJ...792..129N} use a quadratic form, by adding a curvature term, to fit $WISE$ W1 and W2 TF relations.\\

In this paper we present a variation on the NIR TF relation that is specifically designed to be deployed in the ZoA and is applicable to the rest of the sky. We discuss all biases affecting TF surveys in the ZoA in Section 2. In Section 3 we present a re-calibration of the TF relations in the NIR $J$, $H$, and $K_s$ bands using isophotal magnitudes. In the concluding Section 4 we summarize our results and discuss their implications for TF surveys.

\section{Biases in the NIR TF due to extinction}
\subsection{Total magnitude}
Undoubtedly, total magnitudes give the best estimate of a galaxy's total luminosity, and hence, total stellar mass of the galaxy. However, it is difficult to determine reliable total magnitudes for (a) low surface brightness galaxies for which these values can be severely underestimated \citep{2003AJ....125..525J}, or (b) galaxies close to or in the ZoA. Due to the foreground dust extinction and high stellar density in the ZoA, the fainter outer parts of a galaxy are usually not recovered in the extrapolation of the surface brightness profiles to yield total magnitudes \citep*{2010MNRAS.401..924R}. This is particularly severe in the relatively shallow 2MASS as discussed in details by \cite{2002A&A...382..495A} and \cite{2008AJ....136.1866K}. Therefore, the advantage of using isophotal magnitudes over total is twofold.

To quantitatively determine the loss in total magnitudes for ZoA galaxies in 2MASX, we compare the magnitudes of the shallow 2MASX survey with a deeper and more spatially resolved ZoA survey. This is based on a deep NIR imaging conducted with the Japanese InfraRed Survey Facility (IRSF) mounted on a 1.4 m telescope situated at the South African Astronomical Observatory
(SAAO) site in Sutherland, South Africa. Our comparison sample consists of 66 galaxies, observed in both 2MASX \citep{2000AJ....119.2498J} and the IRSF Catalog (\citealt{2011arXiv1107.1096W}; \citealt{2014MNRAS.443...41W}; Said et al. in prep.). The latter is approximately 2 magnitudes deeper in the $K_s$-band, because of its longer exposure time ($10$ min) compared to 2MASX ($8$ seconds) and the higher resolution of 0\farcs 45/pix compared to the 2MASX resolution of 2 $\!\!^{\prime\prime}$/pix. We use the ``total'' extrapolated magnitudes and fiducial ``isophotal'' magnitudes measured in an elliptical aperture defined at the $K_s=20$ mag arcsec$^{-2}$ to construct a comparison between 2MASX and IRSF. The metric used to calculate the offset between 2MASX and ZoA-IRSF magnitudes is

\begin{equation}
\Delta m = m(\text{2MASS})-m(\text{IRSF}).
\end{equation}

\begin{figure}
\begin{center}
\includegraphics[scale=0.35]{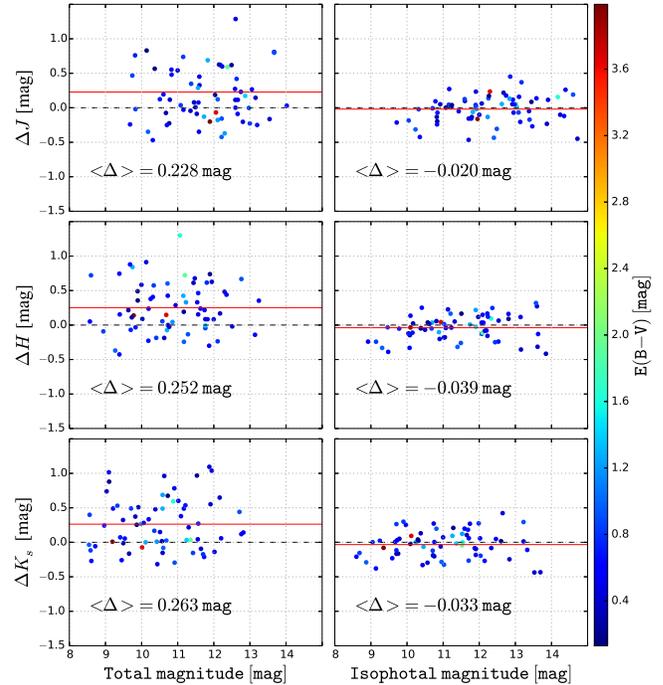}
\caption[Comparison of 2MASX photometry with the $\sim$2 mag deeper and more resolved IRSF photometry of ZoA galaxies]{Comparison of 2MASX photometry with the $\sim$2 mag deeper and more resolved IRSF photometry of ZoA galaxies. Each dot presents galaxy color coded by dust extinction. The left panels display the total magnitudes and the right panels display the isophotal magnitudes with $J$, $H$, and $K_s$-band arranged from top to bottom. In all panels, solid red lines mark the mean offset, and the dashed black lines represent the zero-line. The mean offset is given in the bottom left corner of each plot.}
\label{tfs2}
\end{center}
\end{figure}

Figure \ref{tfs2} shows the offsets between the 2MASS and IRSF for both total (left panel) and isophotal (right panel) magnitudes in the $J$, $H$, and $K_s$-bands (top to bottom). Each dot presents galaxy color coded by dust extinction. It clearly shows that the mean offsets (red line) between 2MASX and IRSF for the ``isophotal'' magnitudes are insignificant while, the offsets are significant for the total magnitude, where the 2MASS estimates are systematically too faint compared to the deeper IRSF. High offsets will produce systematic errors in the measurement of peculiar velocities. The mean offsets for each band is printed in the bottom left corner of each plot. The mean offsets in total magnitudes for $J$, $H$, and $K_s$ bands are 0.23, 0.25, and 0.26 mag respectively. These offsets are large enough to create an artificial flow of about 200 - 600 km s$^{-1}$ in the velocity range of 2000 - 6000 km s$^{-1}$ if the galaxies are located in a similar patch on the sky. The scatter in the offsets of the ``isophotal'' magnitudes is far smaller than that in the ``total'' magnitudes; it is reduced by more than a factor of two. Table \ref{T1} summarizes the mean offsets and the scatter values for each band. 

\begin{table}
\begin{center}
\caption[Comparison between 2MASX and IRSF magnitudes]{Comparison between 2MASX and IRSF magnitudes using the total extrapolated magnitudes and fiducial isophotal magnitudes measured in an elliptical aperture defined at the $K_s=20$ mag arcsec$^{-2}$}
\begin{tabular}{c c c c c}
\hline
\hline
Band & \multicolumn{2}{c}{Total} & \multicolumn{2}{c}{Isophotal}\\
\cline{2-5}
     & $\Delta m$ & $\sigma$ & $\Delta m$ & $\sigma$\\ 
\hline
$J$ & 0.228 & 0.363 & $-0.020$ & 0.155\\
$H$ & 0.252 & 0.293 & $-0.039$ & 0.139\\
$K_s$ & 0.263 & 0.308 & $-0.033$ & 0.159\\
\hline
\end{tabular}
\label{T1}
\end{center}
\end{table} 

This positive offset in total magnitude is due to the improved sensitivity and higher resolution (0\farcs 45/pix) of the IRSF/SIRIUS data, which allows measurement of the surface brightness profile further along the disk compared to 2MASS thus capture more of the host galaxy light. Although minimal, there is a small offset between the 2MASX isophotal magnitudes compared to the IRSF. They are about 0.02 - 0.03 mag brighter rather than fainter. This can be explained by the lower resolution of 2MASS (2 $\!\!^{\prime\prime}$/pix), which does not always resolve stars superimposed on the galaxy resulting in a systematic brightening of the source.

\begin{figure}
\begin{center}
\includegraphics[scale=0.4]{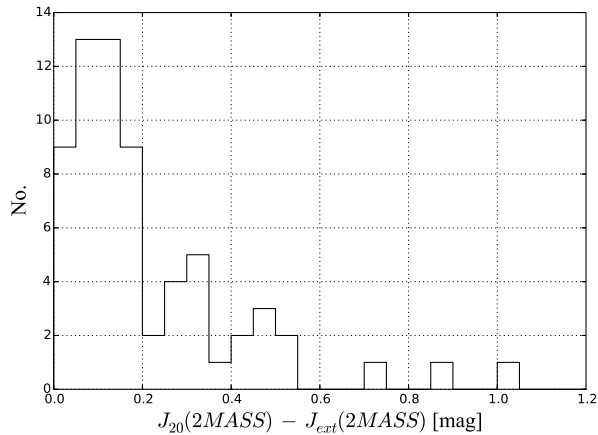}
\caption[Aperture correction]{2MASS aperture correction, $J_{20}-J_{ext}$, of 66 galaxies observed both in 2MASX and IRSF. The mean apperature correction is 0.21 mag.}
\label{aperture_correction}
\end{center}
\end{figure}

Because of these two effects, the loss of low surface brightness flux in 2MASX total magnitudes and extinction which exacerbates this effect, it seems prudent to use isophotal magnitudes, despite the fact that total magnitudes are better tracers than isophotal magnitudes. To test for any systematic bias, we compare the difference in total and isophotal flux. Figure \ref{aperture_correction} presents the aperture correction, $J_{20}-J_{ext}$, of the same 66 galaxies used in Fig. \ref{tfs2}. Figure \ref{aperture_correction} shows that 2MASS isophotal magnitude $J_{20}$ underestimates the total flux by $<J_{20}-J_{ext}>$ = 0.21 mag; in contrast, Fig. \ref{tfs2}, left panel, shows that 2MASS total magnitude underestimates the total flux of 64\% of the sample by $<J_{ext}(\text{2MASS})-J_{ext}(\text{IRSF})>$ = 0.47 mag and overestimates 36\% by 0.19 mag. Thus, any bias introduced by using isophotal magnitude is much smaller than the offsets introduced by using total magnitude in the ZoA.\\

To summarize, when working with 2MASS data for low surface brightness galaxies or galaxies affected by dust, ``isophotal'' magnitudes should preferentially be used over ``total'' magnitudes and most certainly for any TF derived distances. For this reason we derive in Section 3, a calibration of the NIR isophotal TF relation. The method and galaxy sample are exactly the same as these used by \cite{2008AJ....135.1738M} to derive their NIR TF for total magnitude.

\subsection{Inclination}
Spiral galaxies consist of a disk plus a bulge. The disk is almost flat and contains large amounts of dust and young stars, while the bulge is rounder and tends to be dust-poor with older generations. The effect of dust within the galaxy on the derived photometric parameters has been investigated by several authors (e.g., \citealt{1992ApJ...391..617H,1994AJ....107.2036G,2003AJ....126..158M,2004A&A...419..821T,2007MNRAS.379.1022D,
2010MNRAS.404..792M,2013A&A...553A..80P,2013A&A...557A.137P}). In this section we analyze another type of dust-induced change, namely, the change in the shape of the spiral galaxies due to the foreground dust in the Milky Way.\\

The effect of foreground dust extinction on the isophotal radius and magnitude of galaxies (elliptical as well as spirals) has been investigated for optical imaging by \citet{1990A&A...233...16C} and more recently and extensively in the NIR $J$, $H$, and $K_s$ bands by \citet{2010MNRAS.401..924R}. These studies were restricted to analyze the effect of dust on the magnitude and large diameter but not the minor axis, and therefore not the change in the apparent shape (axial ratio) of a galaxy which can vary considerably depending on bulge to disk ratio and intrinsic inclination.\\

Wherever dust obscuration occurs, such an effect should be accounted for. The first reason is the dependence of the TF parameters on inclination. The second reason is to avoid a systematic bias in the sample selections which generally are constrained by inclination uncorrected for absorption effects.

\subsubsection{Inclination-dependent parameters in the TF relation}
In the TF relation two parameters depend on the inclination of a spiral galaxy. The first one is the HI spectrum. To derive the maximum rotational velocity the observed line-width needs to be corrected to edge-on orientation. According to \citet{2007ApJS..172..599S} this is:
\begin{equation}
W=[\frac{w_{50}-\Delta_{s}}{1+z}-\Delta_t]\frac{1}{\sin i}
\end{equation}
where $\Delta_{s}$ and $\Delta_t$ are the instrumental and turbulence corrections, respectively. The inclination $i$ is derived from the 2MASS $J$-band ellipticity $\epsilon_J = 1-(b/a)_J$ via: 
\begin{equation}
\cos^2i = \frac{(1-\epsilon)^2-q_0^2}{1-q_0^2}
\end{equation}
(e.g., \citealt{1997AJ....113...53G}) where $q_0$ is the intrinsic axial ratio of the galaxy ($q_0=0.13$ for Sbc and Sc, and $q_0=0.2$ for other types).  The second set of dependency is with the NIR magnitudes, where we use the equation adopted by a number of authors (e.g., \citealt{1994AJ....107.2036G,1998AJ....115.2264T,2003AJ....126..158M}) to correct for the internal extinction.
Inaccurate inclinations of spiral galaxies will create a systematic bias in both corrected line-width and absolute magnitude, and therefore in the derived peculiar velocities.

\subsubsection{Systematic selection effect}
To minimize the corrections in the line-width, most TF surveys apply a certain lower inclination limit on the axial ratio. The 2MTF project uses only galaxies with $b/a < 0.5$ \citep{2008AJ....135.1738M,2013MNRAS.432.1178H,2014MNRAS.443.1044M}, while in \citet{SAIP_KS} we extend that limit to include galaxies with $b/a < 0.7$. Inaccurate inclinations will not only increase the uncertainty but can be systematic in that galaxies that appear rounder or more inclined will be excluded or included. This effect is dependent on the dust column density along the line-of-sight to the galaxies and as we show, is important for ZoA galaxies. It should therefore be explored how the foreground dust changes the apparent axial ratio. We quantify the effect of the foreground extinction on the number of galaxies in the TF sample through a statistical analysis based on simulated galaxies. The advantage of using artificial galaxies is twofold: firstly it measures the effect on the same galaxy with and without foreground dust; secondly we can test any correction model by applying it to the obscured galaxies and compare their axial ratio to their original intrinsic axial ratio.\\

The IRAF \textit{artdata} tasks \textit{gallist} and \textit{mkobject} were used in this simulation. We created a sample of over 2000 artificial spiral galaxies with axial ratios $b/a \leq 0.5$ to be our primary TF sample. Four other samples have been generated with the same properties except for the magnitude zero point equivalent to adding a dust layer with increasing column density levels which we track accordingly.\\

\begin{figure}
\begin{center}
\includegraphics[scale=0.43]{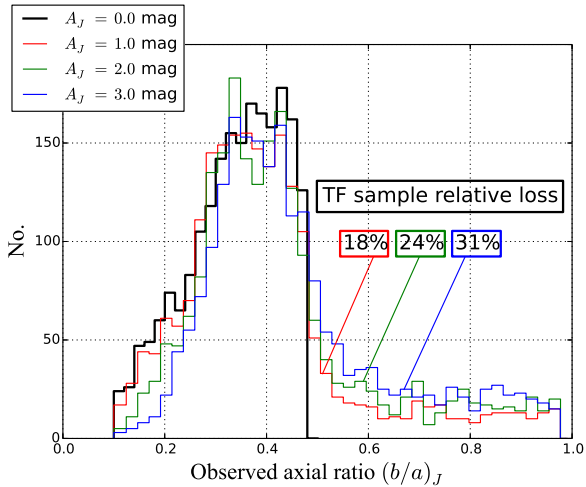}
\caption[The axial ratio distribution of simulated galaxies.]{The axial ratio distribution of simulated galaxies at different levels of dust extinction. The histograms shift to the right, i.e. galaxies become less inclined, and the number of galaxies with $b/a \leq 0.5$ decreases with increasing obscuration level. At dust extinction $A_J$ of  1, 2, and 3 mag, the TF sample relative loss is 18\%, 24\%, and 31\% respectively.}
\label{G_ba_new}
\end{center}
\end{figure}

Figure \ref{G_ba_new} shows the distribution of axes ratios of galaxies with different levels of dust extinction. Note the overall shift of the histograms to the right. The analysis confirms and quantifies our suspicion that galaxies appear rounder with increasing obscuration, which will affect the linewidth and internal extinction correction. Moreover, due to the axial ratio limit of the TF sample, the number of galaxies with $b/a \leq 0.5$ decreases significantly with increasing obscuration level. At dust extinction $A_J$ of  1, 2, and 3 mag, the total number of galaxies in the TF sample decreases by 18\%, 24\%, and 31\% respectively.\\

Figure \ref{G_ba_new} clearly demonstrates how significant this effect can decrease the number of galaxies used in the TF analysis, and therefore bias the survey sample. Future TF surveys based on HI surveys like WALLABY will exclude thousands of galaxies due to the axial ratio limit if this effect is ignored, and this will bias cosmic flow derivations as a function of foreground extinction.\\

To further study this effect, a sample of galaxies drawn from the 2MASS Large Galaxy Atlas (LGA) \citep{2003AJ....125..525J} is used to quantify this effect and test whether a correction model can be applied.

\subsubsection{Quantitative analysis}
In this section we use a sample of 54 spiral galaxies extracted from the 2MASS LGA. This sample of galaxies is selected to cover a wide range in galaxy size,  brightness, morphological types and axial ratios. The sample contains both barred and unbarred galaxies. The galaxies have been artificially dimmed, similar to \citet{1990A&A...233...16C} and \citet{2010MNRAS.401..924R} but for the full 2-dimensional imaging data. We measured the surface brightness profile for both the major and minor axis by using the axial ratio as a free parameter.

\begin{figure}
\begin{center}
\includegraphics[scale=0.45]{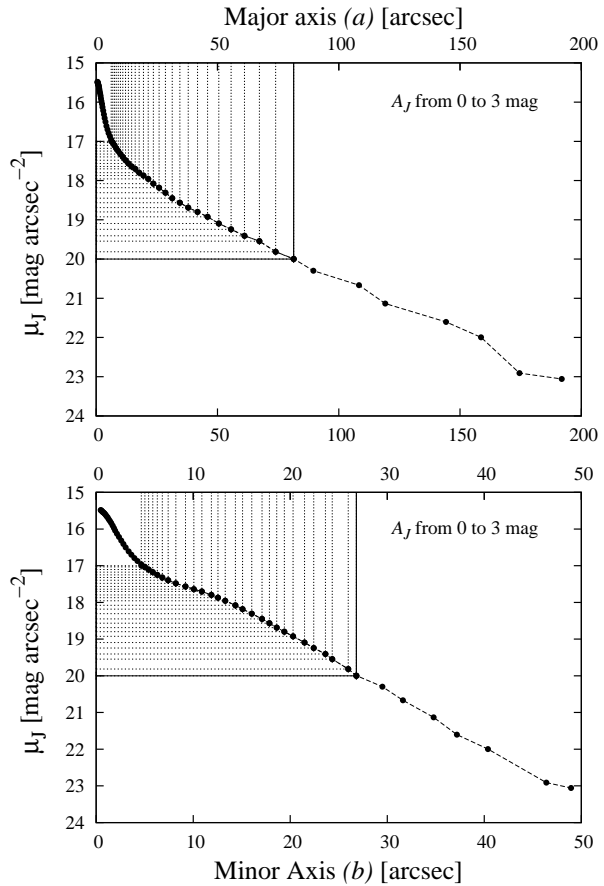}
\caption[Surface brightness profile of NGC1515 (SAB(s)bc)]{Surface brightness profile of NGC1515 (SAB(s)bc). The top panel shows the light profile against the major axis $a$. The bottom panel shows the light profile against the minor axis $b$. Note the different scales for major (top) and minor (bottom) axis. The dotted lines show different levels of dust extinction.}
\label{surf3}
\end{center}
\end{figure}

\begin{figure}
\begin{center}
\includegraphics[scale=0.45]{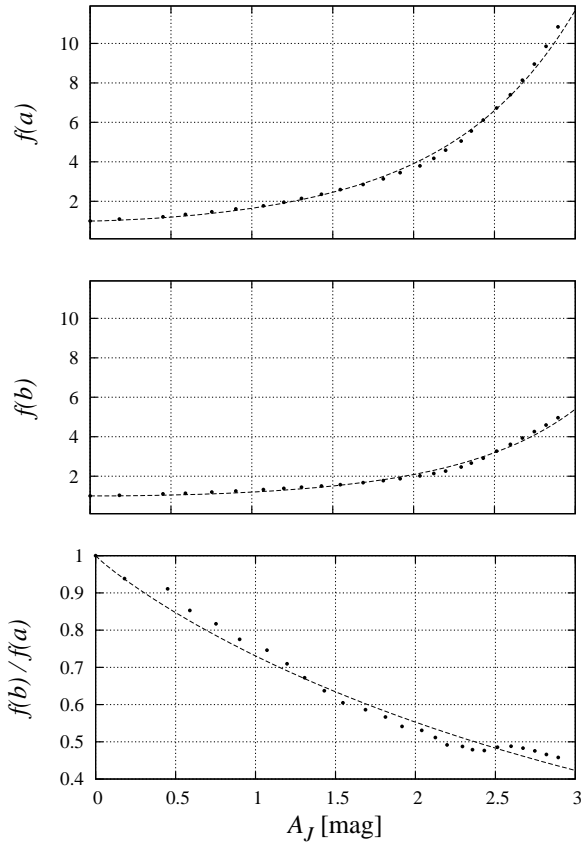}
\caption[NGC1515 galaxy major and minor axis correction due to the foreground extinction]{NGC1515  galaxy major axis correction due to the foreground extinction $f(a)$ in the top panel, the minor axis correction due to the foreground extinction $f(b)$ in the middle panel, and the axial ratio correction due to the foreground extinction $f(b)/f(a)$ in the bottom panel. The dashed lines in all panels show the fit to the data.}
\label{surf4}
\end{center}
\end{figure}

 We now demonstrate the method for the case of NGC1515 (Sbc). It has an intrinsic axial ratio of $b/a=0.34$ in the $J$-band. In Fig. \ref{surf3}, the surface brightness is plotted against the major axis in the top panel and the minor axis in the bottom panel. The projection of the intercept between the solid line and the light profile on the x-axis gives the intrinsic major axis $a^\circ$ (top panel) and minor axis $b^\circ$ (bottom panel). We vary the simulation limits from $A_J = 0.0$ to $A_J = 3.0$ mag. The inward displacements represented by the dotted lines in Fig. \ref{surf3} show the simulated extinction levels $A_J$ in mag. For each level of extinction the major and minor axes have been measured from the intercept of the dotted line and the light profile. We then calculate the ratios of intrinsic versus absorbed axes ratio as a function of extinction.

\begin{subequations}
\begin{eqnarray}
f(a) &=& a^\circ/a, \\
f(b) &=& b^\circ/b,
\end{eqnarray}
\end{subequations}
where $a^\circ$, $a$, $b^\circ$, and $b$ are the intrinsic and absorbed major and minor axes. The functions $f(a)$ and $f(b)$ are plotted for different values of simulated extinction $A_J$ (top and middle panels) in Fig. \ref{surf4}. The bottom panel shows the ratio of $f(b)/f(a)$ against $A_J$, which clearly demonstrates that the axial ratio $(b/a)$ increases with dust extinction (i.e. galaxies become increasingly rounder with increasing extinction). Using the formalism of \citet{1990A&A...233...16C}
\begin{equation}
f(R) = 10^{c(A_\lambda)^{d}}
\end{equation}
where $c$ and $d$ are derived from the data points in Fig. \ref{surf4}.

 The galaxy NGC1515, with an intrinsic axial ratio of $b/a=0.34$ in the $J$-band would appear to have $b/a=0.48$ if seen through 1 mag of extinction, respectively $b/a=0.62$ if seen through 2 mag of extinction. Deep in the ZoA where the extinction level would be very high, this edge-on galaxy would have been  excluded from a typical TF survey analysis.

This procedure was followed for all 54 galaxies in our selected LGA sample and the change in parameters were used to construct an inclination correction model.

We will now investigate the correlations between the change in the axial ratio and the properties of the galaxies, such as, inclination or Hubble type and construct a correction model for inclination. The central surface brightness $\mu_{c}$ and the half-light effective mean surface brightness $\mu_{e}$ are used as indicator of morphological type (see Fig. 18 in \citealt{2003AJ....125..525J}).

Following the recipe given by \citet{2010MNRAS.401..924R} we include the central surface brightness $\mu_{c}$ in the $J$, $H$, and $K_s$-bands to optimize the correction model. The correction equations become
\begin{equation}
f(R,\mu_c) = 10^{c(\mu_c)(A_\lambda)^{d(\mu_c)}},
\end{equation}
where
\begin{equation}
c(\mu_c)=c_0 \exp(\mu_c\times c_1),
\end{equation}
\begin{equation}
d(\mu_c)=d_0 \exp(\mu_c\times d_1).
\end{equation}
The values of $c_0$, $c_1$, $d_0$, and $d_1$ in the $J$, $H$, and $K_s$-bands are the fitting parameters. The mean value for each parameter derived from all galaxies is given in Table \ref{mucpa}. 

\begin{table*}
\begin{minipage}{140mm}
\begin{center}
\caption[The fitting parameters of the correction based on $\mu_c$]{The fitting parameters of the inclination correction based on the central surface brightness $\mu_c$}
\begin{tabular}{l c c c c c c}
\hline
\hline
& \multicolumn{2}{c}{$J$}&\multicolumn{2}{c}{$H$}&\multicolumn{2}{c}{$K_s$}\\
\cline{2-7}
Param. & $b/a\leq0.4$ & $b/a>0.4$ & $b/a\leq0.4$ & $b/a>0.4$ & $b/a\leq0.4$ & $b/a>0.4$\\
\hline
$c_0$ & $-0.0165$ & $-0.0049$ & $-0.0221$ & $-0.0109$& $-0.0328$&  $-0.0254$\\
$c_1$ &  0.1314   & 0.1624    & 0.1161    & 0.1239   & 0.1079   &  0.0762\\
$d_0$ &  3.4042   & 5.0117    & 13.0965   & 0.9383   & 1.1714   &  1.2599\\
$d_1$ & $-0.0510$ & $-0.0742$ & $-0.1387$ & 0.0263   & $-0.0026$&  0.0125\\
\hline
\end{tabular}
\label{mucpa}
\end{center}
\end{minipage}
\end{table*} 

The same method was applied using the half-light effective mean surface brightness $\mu_{e}$. Table \ref{muepa} give the mean value for each parameter.

\begin{table*}
\begin{minipage}{140mm}
\begin{center}
\caption[The fitting parameters of the correction based on $\mu_e$]{The fitting parameters of the inclination correction based on the effective surface brightness $\mu_e$}
\begin{tabular}{l c c c c c c}
\hline
\hline
& \multicolumn{2}{c}{$J$}&\multicolumn{2}{c}{$H$}&\multicolumn{2}{c}{$K_s$}\\
\cline{2-7}
Param. & $b/a\leq0.4$ & $b/a>0.4$ & $b/a\leq0.4$ & $b/a>0.4$ & $b/a\leq0.4$ & $b/a>0.4$\\
\hline
$c_0$ & $-0.0044$ & $-0.0004$ & $-0.0061$ & $-0.0056$& $-0.0100$&  $-0.0537$\\
$c_1$ &  0.1889   & 0.2756    & 0.1754    & 0.1433   & 0.1641   &  0.0220\\
$d_0$ &  9.4482   & 3.2485    & 101.65    & 0.1574   & 8.5084   &  1.9440\\
$d_1$ & $-0.1012$ & $-0.0394$ & $-0.2394$ & 0.1241   & $-0.1186$&  $-0.0144$\\
\hline
\end{tabular}
\label{muepa}
\end{center}
\end{minipage}
\end{table*}

\subsubsection{Applying the correction model}
\begin{figure}
\begin{center}
\includegraphics[scale=0.43]{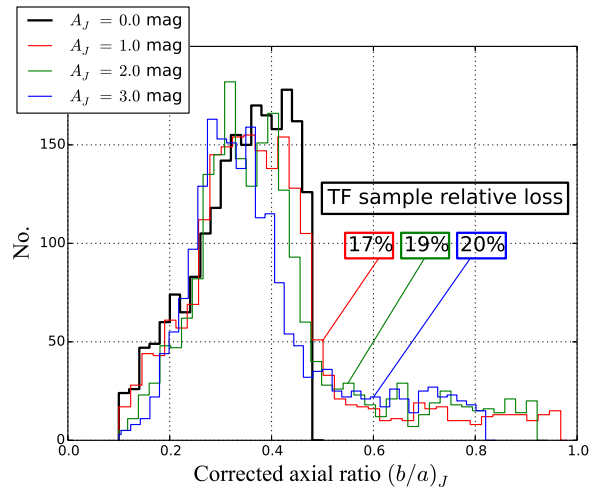}
\caption[The axial ratio distribution of simulated galaxies.]{The axial ratio distribution of simulated galaxies at different levels of dust extinction after inclination correction.  At dust extinction $A_J$ of  1, 2,and 3 mag, the TF sample relative loss is 15\%, 19\%, and 20\% respectively. It may be compared with the uncorrected results in Fig. \ref{G_ba_new}}
\label{G_ba_new_cor}
\end{center}
\end{figure}

To test whether our model reproduces the intrinsic axial ratio from the absorbed value, we compare the corrected values with the intrinsic values (see Fig. \ref{G_ba_new_cor}). We applied the correction based on $\mu_e$ (Eq. 6 and Table \ref{muepa}). 
Figure \ref{G_ba_new_cor} presents the axial ratio distribution of simulated galaxies at different dust extinction level after applying the correction model. At dust extinction $A_J$ of  1, 2, and 3 mag, the TF sample relative loss, after the correction, is 15\%, 19\%, and 20\% respectively. 

The effect of the correction model on the shape of the galaxy is hardly noteable in regions of low extinctions. At extinction level of $A_J\simeq1$, we can recover only 4\% of the TF sample relative loss (see, Figs. \ref{G_ba_new} \& \ref{G_ba_new_cor}).  In regions of higher extinction the improvement in parameters, particularly for the large galaxies seems to be significant. We can recover up to 18\% and 59\% at extinction level of $A_J\simeq2$, and $A_J\simeq3$ mag respectively. In conclusion, applying the correction model is not crucial at high latitudes, nevertheless, it is essential when working in the ZoA where the effect of dust extinction can be sever on the selection of the TF sample.

\section{NIR isophotal TF relation}
In this section we reiterate the advantages of NIR isophotal magnitude when applying the TF relations. The isophotal magnitude, specifically the 20 mag arcsec$^{-2}$ fiducial measurement, is the primary brightness metric for 2MASX \citep{2003AJ....125..525J}. It is uniformly measured across the sky for all galaxies. Total magnitudes are also measured, however they are known to be less reliable and prone to systematic underestimates, notably for low surface brightness galaxies and for confused environments (ZoA). Because of this, many authors prefer using 2MASX isophotal magnitudes instead of total magnitudes to avoid the large uncertainties in the required aperture correction \citep{2000ASPC..218..111B,2001PhDT.......242M,2002A&A...396..431K}. For example, \cite{2001PhDT.......242M} used both total and isophotal magnitude to derive TF relations. The slope of his isophotal magnitude relations in $H$, and $K_s$ is shallower than the total magnitude relation. He tested these isophotal relations against the $I$-band total magnitude and concludes that both isophotal $H$ and $K_s$ relations and $I$ band total relation provide distance estimates with similar precision. Another TF study by \cite{2000ASPC..218..111B} tests a variety of magnitudes provided by 2MASX and found that the scatter was lowest for the isophotal $K_s$ magnitude. These studies agree with our findings in Section 2 that there are clear advantages of working with isophotal magnitudes over total magnitudes and that using isophotal TF relation in and out of the ZoA does not introduce any potential bias in the derived distance and peculiar velocity.
\subsection{The TF sample}
Two observational parameters describe the TF relation. The distance-independent rotation width is obtained from spectral data, and the imaging data provide all photometric quantities of interest. Different authors prefer different quantities, such as, the 20\% line-width versus the 50\% line-width, and isophotal versus total magnitudes. A variety of methods have been used over the years to derive $w_{50}$ \citep{2004AJ....128...16K,2005ApJS..160..149S,2013MNRAS.432.1178H}. These different algorithms can have a serious effect on the accuracy of the value of $w_{50}$ if the S/N of the spectra are low. Consequently, TF samples only include galaxies with high S/N spectra, so that the noise will not affect the width measurement and the difference between different algorithms becomes negligible.\\

The calibration sample of galaxies is the same as that used in \citet{2008AJ....135.1738M}\footnote{We thank 
Karen Masters for making that available to use}. Their rotation-widths were obtained from either the Cornell HI digital archive \citep{2005ApJS..160..149S} or optical rotation curves \citep{2005AJ....130.1037C}. The photometric quantities of the calibration sample are a mix of $J$, $H$, and $K_s$-bands quantities from 2MASX and $I$-band quantities from \citet{2006ApJ...653..861M}. We cross-matched the \citet{2008AJ....135.1738M} calibration sample with the 2MASX catalog and extracted the ``isophotal'' magnitudes for all of them.

\subsubsection{Rotation widths}
The rotation widths are drawn completely from the \citet{2008AJ....135.1738M} calibration sample which is available online\footnote{{http://www.icg.port.ac.uk/$\sim$mastersk/TFdata.html}}, see \citealt{2005ApJS..160..149S} for details.

\subsubsection{Photometry}
All photometric quantities are derived from the 2MASX catalog. Their properties are described online in Cutri et al. (2006)\footnote{{http://www.ipac.caltech.edu/2mass/releases/allsky/doc/}}. The photometric quantities of the TF calibration are:\\

1. Isophotal magnitude: We use the $J$, $H$, and $K_s$-bands fiducial isophotal magnitudes measured at 20.0 mag arcsec$^{-2}$ in $K_s$ band, which is roughly equal to the 1 $\sigma$ background noise level \citep{2003AJ....125..525J}.

2. $J$-band axial ratio:  The $J$-band axial ratio $(a/b)_J$ fit to the 3 $\sigma$ isophote is used (the isophote corresponds to a surface brightness $\sim$ 3 times the background noise).

3. Central surface brightness $\mu_c$: The central surface brightness $\mu_c$ in the $J$, $H$, and $K_s$-bands obtained from the 2MASS XSC. This parameter is necessary for the derivation of the Galactic extinction corrections (Section 4.1.3).

4. The colour reddening $E(B-V)$: Galactic extinction $A_J$, $A_H$, and $A_{K_s}$ are approximated by the colour reddening $E(B-V)$ from the DIRBE dust map \citep{1998ApJ...500..525S}. Dust extinction of the calibration sample are very low and the difference between \cite{1998ApJ...500..525S} values and its updated values from \cite{2011ApJ...737..103S} have no effect on our calibration. However for high extinction regions we use the \cite{2011ApJ...737..103S} values of 0.87 times \cite{1998ApJ...500..525S} values which is in excellent agreement with the independent derived correction in the ZoA by \cite{2007A&A...466..481S} and more recently by \cite{2014MNRAS.443...41W}.

\subsection{Photometric corrections}   
To construct a global TF relation, the photometric quantities from the 2MASX must be corrected for the cosmological redshift, internal extinction, and Galactic extinction. The corrected absolute magnitude, derived from the observed apparent magnitude is calculated as follows\footnote{The sign error in the application of the internal extinction and k-corrections to magnitudes used in \cite{2008AJ....135.1738M} and corrected in \cite{2014AJ....147..124M} has been accounted for.}:
\begin{equation}
M_{corr} - 5 \log h = m_{obs} - A_X - I_X - k_X - 5 \log v_{CMB} - 15,
\end{equation}
where $A_X$, $I_X$, and $k_X$ are a correction for foreground extinction due to the dust in the Milky Way, a correction for extinction internal to the galaxy itself, and a cosmological $k$-correction respectively. Because these three corrections are wavelength-dependent, the index $X$ refers to the wavelength band. For the $k$-correction we used the same procedure used by \cite{2008AJ....135.1738M}. In the next two subsections we describe the methods used to correct for internal and Galactic extinction as we deviate slightly from the \cite{2008AJ....135.1738M} procedure.

\subsubsection{Internal extinction}
Dust extinction internal to galaxies themselves is a challenging quantity to estimate. Since the early work of \citet{1958MeLu2.136....1H} it has been widely assumed that spiral galaxies are mostly transparent.
Later studies suggest that spiral galaxies are optically thick \citep{1990Natur.346..153V}. Again, this study received some criticism (e.g. \citealt{1991Natur.353..515B}). The best study to date \citep{2009AJ....137.3000H}, uses  occulting galaxy pairs to show that dust distributions vary, creating both optically thin and thick regions.\\

\begin{figure}
\begin{center}
\includegraphics[scale=0.4]{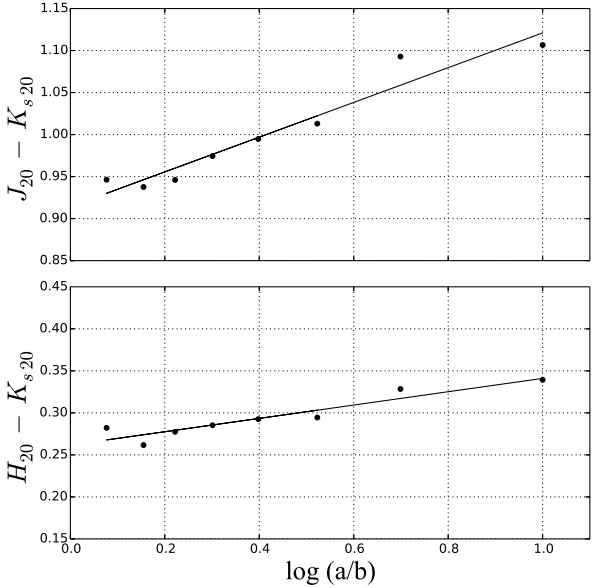}
\caption[]{NIR colour, $J_{20}-{K_s}_{20}$ and $H_{20}-{K_s}_{20}$, dependence on inclination of the calibration sample (888 galaxies) binned in inclination groups. The slopes of Eq. \ref{int_ext_equ} are 0.21 for $J_{20}-{K_s}_{20}$ and 0.08 for $H_{20}-{K_s}_{20}$. We assume that the $K_s$-band isophotal magnitude is internal extinction free (i.e., $\gamma_{K_s}=0$)}
\label{int_ext_corr_iso}
\end{center}
\end{figure}

This problem has been investigated by several authors from different aspects. Most correction models for the internal extinction have been constructed as a function of inclination. In the TF relation, inclined galaxies are preferentially used, but the magnitudes of inclined galaxies are more affected by dust. In our analysis we adopt the empirical relation of internal extinction described in \cite{1994AJ....107.2036G,1998AJ....115.2264T,2003AJ....126..158M}, which is dependent on axial ratio and waveband, as follows:
\begin{equation}
I_{\lambda} = \gamma \text{ log} (a/b).
\label{int_ext_equ}
\end{equation}
The $J$-band axial ratio $(b/a)_J$ is the best tracer of the inclination, as it suffers less than the $H$, and $K_s$ -bands from the effect of the bulge population. The axial ratio needs to be corrected for seeing, which is parameterized as 
\begin{equation}
(a/b)_{corr} = (a/b)_{obs}(1-0.02x+0.21x^{2}-0.01x^{3}),
\end{equation}
where $x=$ FWHM$(\frac{a/b}{r_{20}})$, and FWHM = 2\farcs5 (the full-width half-maximum of the seeing disc). \citet{2003AJ....126..158M} provide different values of $\gamma$ in $J$, $H$, and $K_s$-bands from different statistical tests based on total magnitudes. We deviate slightly in our approach because applying correction for total magnitudes will results in an over-correction in isophotal magnitudes. To investigate the internal extinction in $J$, and $H$ bands, we use the calibration sample, 888 galaxies, to test the colour gradients (measured at 20.0 mag arcsec$^{-2}$ in $K_s$ band) with inclination. We assume that the internal extinction in the $K_s$-band isophotal magnitude is negligible (i.e., $\gamma_{K_s}=0$); the effect of internal extinction is indeed very small in $K_s$ band even in total magnitudes \citep{1998AJ....115.2264T,2003AJ....126..158M}. Figure \ref{int_ext_corr_iso} shows how NIR colours, $J_{20}-{K_s}_{20}$ and $H_{20}-{K_s}_{20}$, change with inclination.

In this Fig. we used the calibration sample, 888 galaxies, binned in inclination groups. The slopes of Eq. \ref{int_ext_equ} are:

\begin{subequations}
\begin{eqnarray}
\gamma_{J_{20}-{K_s}_{20}} &=& \gamma_{J_{20}}-\gamma_{{K_s}_{20}} \nonumber \\
 &=& \gamma_{J_{20}} \nonumber \\ 
 &=& 0.21\pm0.03,\\
\gamma_{H_{20}-{K_s}_{20}} &=& \gamma_{H_{20}}-\gamma_{{K_s}_{20}} \nonumber \\
 &=& \gamma_{H_{20}} \nonumber \\
 &=& 0.08\pm0.02.
\end{eqnarray}
\end{subequations}

These values of $\gamma$ mean that the effect of internal extinction is very small in the isophotal magnitudes compared to total magnitudes because most of the dust are in the disk.

\subsubsection{Galactic  foreground extinction}
Galaxies appear smaller and fainter due to the dust extinction in the Milky Way. Therefore a correction to the isophotal magnitude is needed to account for the loss of light due to this dimming.  \citet*{2010MNRAS.401..924R} present two methods to apply the isophotal correction: ($i$) the average correction method which is a very direct application, and ($ii$) the more optimized correction method which is more accurate than the average correction but requires a fit to the light profile of the galaxy to determine either the disc central surface brightness, $\mu_0$, or the combined disc plus bulge central surface brightness, $\mu_c$. The optimized correction was used here, because the central surface brightness is available in the 2MASX.

\subsection{Bias corrections}
Many TF samples, including the one we use here, contain a broad range of spiral types. A single relation may not be appropriate for all spiral types as expressed already by Tully \& Fisher themselves. \citet{2008AJ....135.1738M} found type dependences with their sample, and our analysis of this effect based on the ``isophotal'' magnitude yields similar trends. The earlier types have a shallower slope than later types. We correct to an Sc type (less bulge, more disk component) relation. In the $J$-band we use
\medskip
\begin{subequations}
\begin{eqnarray}
\Delta M_{Sa} &=& 0.27 - 2.46 (\log W - 2.5),\\
\Delta M_{Sb} &=& 0.15 - 1.14 (\log W - 2.5).
\end{eqnarray}
\end{subequations}
In the $H$-band we use 
\begin{subequations}
\begin{eqnarray}
\Delta M_{Sa} &=& 0.22 - 2.81 (\log W - 2.5),\\
\Delta M_{Sb} &=& 0.15 - 1.16 (\log W - 2.5).
\end{eqnarray}
\end{subequations}
In the $K_s$-band we use 
\begin{subequations}
\begin{eqnarray}
\Delta M_{Sa} &=& 0.11 - 3.51 (\log W - 2.5),\\
\Delta M_{Sb} &=& 0.13 - 1.44 (\log W - 2.5).
\end{eqnarray}
\end{subequations}
These values are derived using the isophotal photometry for each sample before any bias corrections.\\

Due to the broad range in surface brightness and colour of the galaxy sample, further bias corrections including Incompleteness bias, Cluster size bias and Cluster peculiar velocity have to be applied. We used the values derived by \citet{2008AJ....135.1738M} to correct for these biases. Any difference in the correction would be incredibly small and much lower than any of the other corrections.

\subsection{Calibration and scatter}
Different fitting procedures have been used to derive the final parameters of the TF relation. The inverse TF was suggested by \citet{1986gddu.work...65K} to overcome the Malmquist bias. \citet{1997AJ....113...53G} used the minimization of $\chi^2$ method to determine the direct, inverse, and bivariate forms of the linear TF relation. In \citet{2007AJ....134..945P}, the maximum likelihood method was used to estimate the slope $a$, intercept $b$, and intrinsic scatter $\epsilon_{int}$ of the TF relation.
Here we are interested in the bivariate fit case, where errors in both $x$ and $y$ are taken into consideration.
We therefore use the linear form: 
\begin{equation}
y(x)=a_{bi}+b_{bi}x,
\end{equation}
via minimization of $\chi^2$
\begin{equation}
\chi^2=\sum_{i=1}^N[\frac{y_i-y(x_i;a_{bi},b_{bi})}{\epsilon_i}]^2.
\end{equation}
The error used in the computation is defined as $\epsilon_i=[(\epsilon_{x,i}b_{bi})^2+\epsilon_{y,i}^2+\epsilon_{int}^2+\text{cov}_{xy}]^{1/2}$, where $\text{cov}_{xy}$ represents the covariance between the errors in $x$ and $y$ which is not significant in our case.\\
After applying all of the above detailed bias corrections, and applying the bivariate fitting mechanism to the \cite{2008AJ....135.1738M} 888 calibrator galaxies, our final isophotal TF relations have the form:
\begin{subequations}
\begin{eqnarray}
M_J-5 \log h &=& -20.951-9.261(\log W-2.5),\\
M_H-5 \log h &=& -21.682-9.288(\log W-2.5),\\
M_{K_s}-5 \log h &=& -21.861-10.369(\log W-2.5).
\end{eqnarray}
\end{subequations}

\begin{figure}
\begin{center}
\includegraphics[scale=0.45]{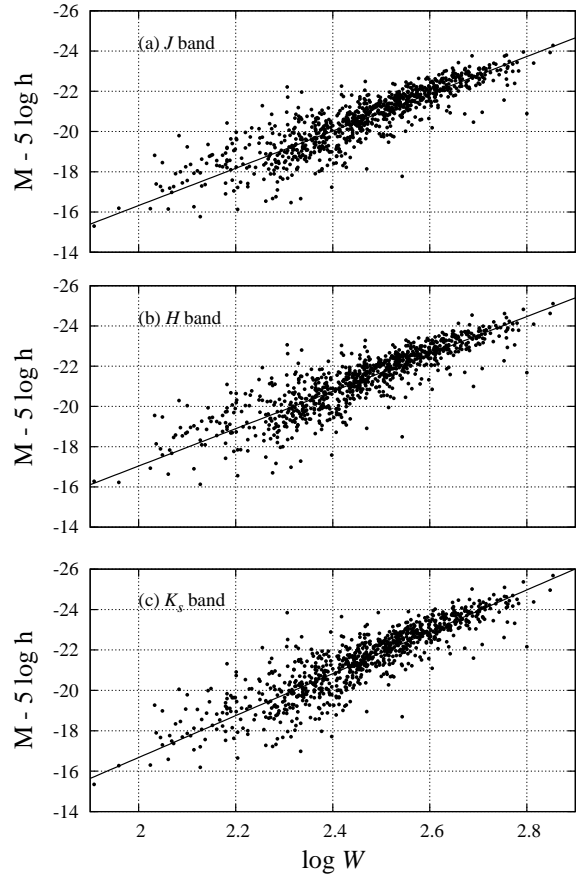}
\caption[Isophotal TF relation]{Isophotal TF relation for the (a) $J$-band, (b) $H$-band, and (c) $K_s$-band. The solid line shows the respective bivariate fit to the data.}
\label{TF2}
\end{center}
\end{figure}

Table \ref{T4} presents the fitted parameters derived from both isophotal magnitude (this work) and total magnitude from \citet{2014AJ....147..124M}.
Figure \ref{TF2} shows the final isophotal TF relations for the $J$, $H$, and $K_s$-band from top to bottom.
\begin{table*}
\begin{minipage}{140mm}
\begin{center}
\caption[Parameters of bivariate fit after corrections]{Parameters of bivariate fit after all corrections derived from both isophotal (this work) and total magnitude \citep{2014AJ....147..124M}.}
\begin{tabular}{ c  c c c c  c c c c }
 \hline
 \hline
Band & \multicolumn{4}{c}{This work} & \multicolumn{4}{c}{\citet{2014AJ....147..124M}}\\
\cline{2-9}
 &$a_{iso}$ & $b_{iso}$ & $\epsilon_a$ & $\epsilon_b$ & $a_{tot}$ & $b_{tot}$ & $\epsilon_a$ & $\epsilon_b$\\
\hline
$J$-band &$-$20.951 & $-$9.261 & 0.018 & 0.114  & $-$21.370 & $-$10.61 & 0.018 & 0.12\\
$H$-band &$-$21.682 & $-$9.288 & 0.018 & 0.115  & $-$21.951 & $-$10.65 & 0.017 & 0.11 \\
$K_s$-band &$-$21.861 & $-$10.369 & 0.018 & 0.120  & $-$22.188 & $-$10.74 & 0.015 & 0.10\\
\hline

\end{tabular}
\label{T4}
\end{center}
\end{minipage}

\end{table*}

\begin{figure}
\begin{center}
\includegraphics[scale=0.4,angle=270]{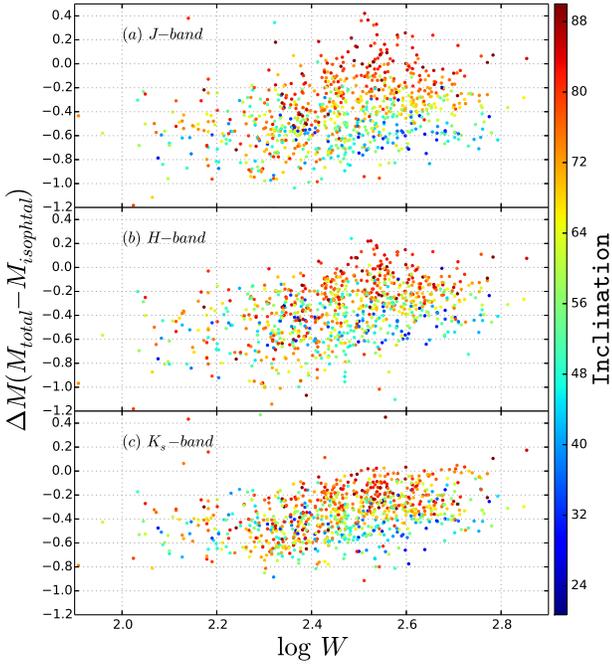}
\caption[]{Total versus isophotal magnitude for the template galaxies as a function of line-width. Galaxies are color coded by its inclination.}
\label{dm_jhk}
\end{center}
\end{figure}

The slope for the isophotal TF relations in $J$ and $H$ bands are shallower than the total TF but similar in the $K_s$ band. This can be explained by Fig. \ref{dm_jhk}, which shows the difference between isophotal and total TF relation as a function of line-width and inclination.

Each dot in Fig. \ref{dm_jhk} presents a galaxy from the TF calibration sample colour-coded by its inclination. The $K_s$ band has the tightest scatter in Fig. \ref{dm_jhk}. This is not a surprising result given that the aperture correction in the $K_s$ band is lowest compared to the $J$ and $H$ bands (see also Fig. 7 in \citealt{2003AJ....125..525J}). The deviation increases toward small and less inclined galaxies whose surface brightness profiles are not well determined. This further supports the preference of using isophotal magnitudes as the total magnitude for those galaxies become unreliable.

\begin{figure}
\begin{center}
\includegraphics[scale=0.45]{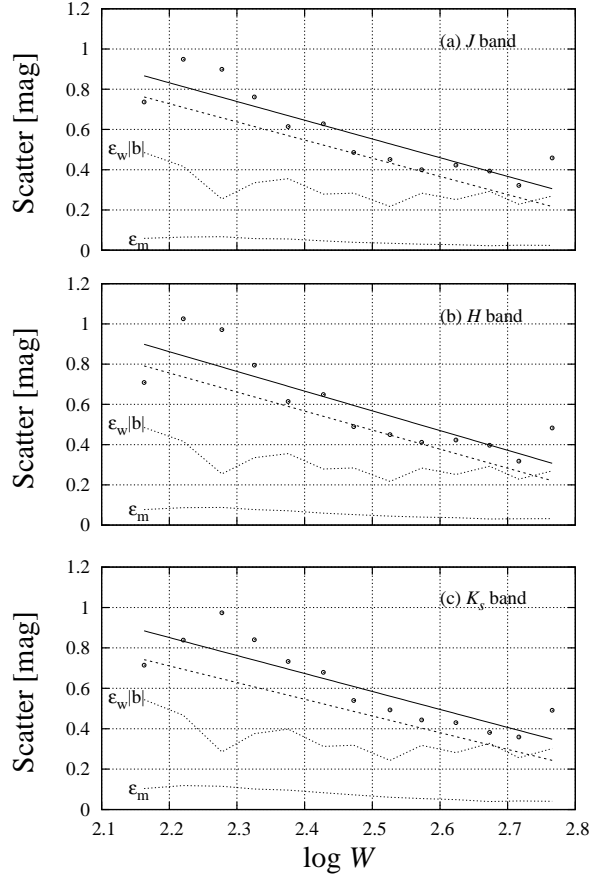}
\caption[Scatter in the (a)$J$-band, (b)$H$-band, and (c)$K_s$-band]{Scatter in the (a)$J$-band, (b)$H$-band, and (c)$K_s$-band. The circles present the total scatter averaged within bins in velocity width. The linear fit to the total scatter is shown as a solid line and the dashed line shows the linear fit to the intrinsic scatter. The dotted lines present the error on the isophotal magnitude and the error on the rotation width multiplied by the slope}
\label{TFscatter1}
\end{center}
\end{figure}

Figure \ref{TF2} also reveals an increasing scatter in the TF relation with decreasing rotation width. Using the relation used in \citet{1997AJ....113...53G} to parameterize the scatter
\begin{equation}
\sigma=a+b(\log W-2.5),
\end{equation}
the total scatter is calculated, and displayed in Fig. \ref{TFscatter1} for the $J$, $H$, and $K_s$-bands as a function of the line-width.\\

The error on the isophotal magnitude (least source of error), as well as the error on the rotation width multiplied by the slope of the TF relation (expressed in magnitudes) are also displayed in Fig. \ref{TFscatter1}. A linear fit to the intrinsic scatter gives
\begin{subequations}
\begin{eqnarray}
\epsilon_{int,J} &=& 0.46-0.90(\log W-2.5),\\
\epsilon_{int,H} &=& 0.47-0.94(\log W-2.5),\\
\epsilon_{int,K_s}&=&0.46-0.83(\log W-2.5).
\end{eqnarray}
\label{isointr}
\end{subequations}

The intrinsic scatter, in magnitude units, is a crucial parameter to understand the errors on both distances and peculiar velocities derived from the TF relation. The intrinsic scatter in both the ``isophotal'' and ``total'' methods are nearly identical, because the calibration sample of galaxies is hardly affected by foreground dust of the Milky Way. 

Comparing equations \ref{isointr} with \citet{2008AJ....135.1738M} results shows that the intrinsic scatter is smaller in the total magnitude relation for the larger galaxies ($\log W > 2.5$), while for smaller galaxies ($\log W < 2.5$) the ``isophotal'' magnitude relation shows less scatter. 

\section{Conclusions}
We study the effect of foreground dust and source confusion on galaxy photometry in the context of employing the TF method in the Zone of Avoidance. Two different methods using independent samples of galaxies were used to quantify this effect. Both methods confirm and quantify that galaxies appear rounder with increasing obscuration level leading to a substantial loss of galaxies in inclination-constrained TF samples deep in the ZoA.
Correction models are proposed based on 54 spiral galaxies from the 2MASS LGA with different morphological types and intrinsic axial ratios. These correction models have been tested and show its applicability to reproduce the intrinsic axial ratio from the observed value for large galaxies up to extinction level of about $A_J\simeq3$ mag, and recover a fair fraction of galaxies that otherwise would fall out of an uncorrected inclination limited galaxy sample.

We present the recalibration of the Tully-Fisher relation for isophotal magnitudes in the NIR $J$, $H$, and $K_s$-bands. This calibration sample of 888 galaxies is the same as the one used for the 2MTF project for total magnitudes. 
No significant change in the isophotal TF relation scatter was found in comparison to the scatter in the total TF relation because the calibration sample is minimally affected by the dust of the Milky Way. However, this does not hold for low surface brightness galaxies or galaxies obscured by dust where isophotal magnitudes are more robust. 

The isophotal NIR TF relation has been applied in a pilot project by \cite{SAIP_KS} to a preliminary sample of HI detected galaxies in the ZoA which had deep NIR photometry \citep{2014MNRAS.443...41W} and found to be quite promising.
Considerable improvement of the scatter and the systematic offset of the derived peculiar velocities in the ZoA is achieved when the isophotal TF relation is applied in comparison to the traditional total magnitude method. The offsets of about 0.2 - 0.3 mag which may create an artificial flow of about 200 - 600 km s$^{-1}$ in the velocity range of 2000 - 6000 km s$^{-1}$, have been reduced to about 0.02 - 0.03 mag in $J$, $H$, and $K_s$ bands. Combining our data with that from 2MTF will provide more complete all-sky peculiar velocity survey. Using the isophotal TF relation and accounting for the effect of extinction on inclination will lead to an improved TF analyses at lower latitudes. It will allow the extension of cosmic flow derivations deeper into the ZoA compared to current surveys and, moreover, improve on the results for galaxies with $|b| \leq 5^\circ$ that are already affected by dust, or any other regions of high extinction in the sky.

\section*{Acknowledgments}
This work is based upon research supported by the South African National Research Foundation and Department of Science and Technology. The authors thank Wendy Williams for communicating her data in the ZoA. The authors are very grateful to Dr. Maciej Bilicki for his reading and comments on the paper and to Dr. Michelle Cluver for many discussions about photometry and luminosity functions. We acknowledge the HIZOA survey team for early access to the data. And foremost, we thank the referee Dr. Karen Masters for very valuable comments. This publication makes use of data products from the Two Micron All Sky Survey, which is a joint project of the University of Massachusetts and the Infrared Processing and Analysis Center, funded by the National Aeronautics and Space Administration and the National Science Foundation.

\bibliographystyle{mn2e.bst}
\bibliography{774}

\label{lastpage}

\bsp

\end{document}